\documentclass[twocolumn]{jpsj2}

\title{
Nonuniform Spin Triplet Superconductivity due to Antisymmetric 
Spin-Orbit Coupling in Noncentrosymmetric Superconductor CePt$_3$Si}

\author{Hiroko \textsc{Tanaka}\thanks{E-mail address: ri05i022@stkt.u-hyogo.ac.jp},
Hirono \textsc{Kaneyasu}\thanks{Present address: Theoretische Physik, ETH-Honggerberg, CH-8093, Zurich, Switzerland}
and Yasumasa \textsc{Hasegawa}
}

\inst{Department of Material Science, University of Hyogo,\\
 3-2-1 Kouto, Kamigori-cho, Hyogo, Akou-gun 678-1297, Japan}

\recdate{\today}

\abst{
We show that
the nonuniform state (Fulde-Ferrel-Larkin-Ovchinnikov (FFLO) state) 
of the spin triplet superconductivity in noncentrosymmetric systems
is stabilized by antisymmetric spin-orbit coupling
  even if the magnetic field is absent. 
The transition temperature of the spin triplet superconductivity is reduced by  
 the antisymmetric spin-orbit coupling in general. 
This pair breaking effect is shown to be 
 similar to the Pauli pair breaking effect due to magnetic field
 for the spin singlet superconductivity,
in which FFLO state is stabilized near the Pauli limit 
(or Chandrasekhar-Clogston limit) of external magnetic field.
Since there are gapless excitations in nonuniform superconducting state,  some  physical quantities
such as specific heat and penetration depth should obey the power low temperature-dependences.
We discuss the possibility of the realization of nonuniform state in CePt$_3$Si.

}

\kword{
CePt$_3$Si, heavy fermion superconductivity, spin-orbit coupling,
FFLO, spin triplet superconductivity
}
\begin{document}
\maketitle

\section{Introduction}
Recently, superconductivity in the system without inversion center attracts much interest.
The first heavy fermion superconductor
without a center of inversion has been observed in CePt$_3$Si,\cite{Bauer}
and several pressure-induced superconductivity in the systems without inversion
center have been reported, 
including CeRhSi$_3$\cite{Kimura}, CeIrSi$_3$\cite{Sugitani} and UIr\cite{Akazawa}.
%
In CePt$_3$Si 
 antiferromagnetic order exists below $T_{N}= 2.2$K 
followed by superconductivity below  $T_{c}= 0.75$K.
The upper critical field $H_{c2}(\sim  5T)$ exceeds the Pauli paramagnetic
limiting field (or Chandrasekhar-Clogston limit)\cite{Bauer}, 
which is thought to be an evidence of the spin triplet pairing.
Recent experiments on Knight shift\cite{Ueda,Yogi2006} also indicate the spin triplet superconductivity.
%
%

On the other hand it has been believed that
inversion symmetry is required for the spin triplet superconductivity\cite{Anderson}.
Frigeri et al.\cite{Frigeri} have studied the spin triplet superconductivity in the 
system without inversion symmetry in detail. 
They took the Rashba type spin-orbit coupling, 
\begin{equation}
  H_p = \alpha \sum_{\mathbf{k}, s,s'} \mathbf{g}_{\mathbf{k}} \cdot \vec{\sigma}_{s,s'}
   c_{\mathbf{k},s}^{\dagger} c_{\mathbf{k},s'} ,
\label{spin-orbit}
\end{equation}
where $c^{\dagger }_{\mathbf{k}s}$ and $c_{\mathbf{k}s}$ are 
creation and annihilation operators of electrons with momentum $\mathbf{k}$ 
and spin $s$, respectively,
and $\mathbf{g}_{\mathbf{k}}=(-k_y, k_x,0)$
as a model for CePt$_3$Si, the point group of which is $C_{4v}$.
%
They obtained that the spin triplet pairing is not affected by this spin-orbit coupling, 
if the $\mathbf{d}$-vector of the spin triplet pairing
 satisfies $\mathbf{d}(\mathbf{k}) \parallel \mathbf{g}_{\mathbf{k}}$,
while the transition temperature of the triplet superconductivity is 
reduced if  $\mathbf{d}(\mathbf{k})
\nparallel  \mathbf{g}_{\mathbf{k}}$.

Another interesting property of superconductivity in CePt$_3$Si is the existence of the line nodes
in the energy gap indicated by the temperature dependence of
penetration depth\cite{bonalde} and thermal conductivity\cite{izawa}. 
Since the spin triplet state with $\mathbf{d}(\mathbf{k}) \parallel \mathbf{g}_{\mathbf{k}}$
proposed by Frigeri et al.\cite{Frigeri}
has points nodes at $k_x=k_y=0$ and no line nodes, this state is not consistent with the experiments. 
In order to explain the existence of the line nodes, the mixing  state of triplet and singlet 
pairings has been studied\cite{hayashi1,hayashi2}. The strong spin-orbit coupling and
the energy dependence of the density of states (particle-hole asymmetry) have to be assumed
to obtain the mixed state of spin singlet and spin triplet.

We propose another possibility,
nonuniform spin triplet state,  to explain the experiments in CePt$_3$Si. 
We will show that there is a similarity between the pair-breaking effect due to 
the anisotropic spin-orbit coupling for the triplet superconductivity with $\mathbf{d}(\mathbf{k})
\nparallel \mathbf{g}_{\mathbf{k}}$ and the Pauli pair breaking effect due to the magnetic field
for the spin singlet superconductivity. 
In the latter case 
 Fulde and Ferrel\cite{Fulde}, and Larkin and Ovchinnikov\cite{Larkin}
have shown that the 
Pauli pair breaking gives rise to a new pairing state between
electrons $(\mathbf{k}+\frac{\mathbf{q}}{2},\uparrow )$ and 
$(-\mathbf{k}+\frac{\mathbf{q}}{2},\downarrow )$ 
on the exchange-split parts of the Fermi surface due to Zeeman effect
of magnetic field, if the external magnetic field is close to the
 Pauli limit. 
This state is called FFLO state
and it has been observed in CeCoIn$_5$\cite{Kakuyanagi}, recently.
In this paper we show that the similar pairing state to FFLO state is possible
for the spin triplet superconductivity with $\mathbf{d}(\mathbf{k})
\nparallel  \mathbf{g}_{\mathbf{k}}$ even if the magnetic field is absent.
In the FFLO state without magnetic field the order parameter is not uniform and the absolute value
of the order parameter is zero in parallel planes with period $\frac{2\pi}{|\mathbf{q}|}$
 in real space. Then
some physical properties obey the power-low temperature dependences.

\section{Model and  order parameters}
We adopt a single-band model with electron band energy $\xi _{\mathbf{k}}$
measured relative to the Fermi energy, which is assumed to be spherical when there are no interactions. 
We use the weak coupling approach, taking the pairing interaction  as 
\begin{align}
H_V 
 = \frac{1}{2}\sum_{\mathbf{k},\mathbf{k}',\mathbf{q}}\sum_{s,s'}V_{\mathbf{kk}'}
c_{\mathbf{k}+\frac{\mathbf{q}}{2},s}^{\dagger }
c_{-\mathbf{k}+\frac{\mathbf{q}}{2},s'}^{\dagger }
c_{-\mathbf{k}'+\frac{\mathbf{q}}{2},s'}
c_{\mathbf{k}'+\frac{\mathbf{q}}{2},s}.
\end{align}
%
 We assume that the interaction is finite and attractive close to
the Fermi energy with the cutoff energy $\epsilon _c$ and
 that the pairing interaction does not depend on the spin and 
the total momentum of the pair, $\mathbf{q}$.
%
%
%
This Hamiltonian satisfies time reversal and inversion symmetry.
The absence of inversion symmetry is introduced by spin-orbit coupling term, 
$H_p$ (eq.~(\ref{spin-orbit})),
with
\begin{equation}
 \mathbf{g}_{\mathbf{k}} = \sqrt{\frac{3}{2}} \frac{1}{k_F} (-k_y,k_x,0),
\end{equation} 
 as studied by Frigeri et al.\cite{Frigeri}.
For simplicity, we set ${\langle {\mid \mathbf{g_k} \mid}^{2}\rangle}_{\mathbf{k}}=1$ 
where $\langle \cdots \rangle_{\mathbf{k}}$ 
denotes an average over the Fermi surface.
Then, the single-particle Hamiltonian is 
\begin{equation}
H_0=\sum_{\mathbf{k},s,s'}[\xi _{\mathbf{k}}+\alpha \mathbf{g_k}\cdot \vec{\sigma }]_{ss'}
c_{\mathbf{k}s}^{\dagger }c_{\mathbf{k}s'}.
\label{H0}
\end{equation}

The normal state Green's function is obtained in the $2\times 2$ matrix form as
\begin{equation}
 G^{0}(\mathbf{k}, i\omega _{n})=
 G_{+}(\mathbf{k}, i\omega _{n})\sigma _{0}+
 (\hat{\mathbf{g}}_{\mathbf{k}}\cdot \vec{\sigma} )
 G_{-}(\mathbf{k}, i\omega _{n}).
\end{equation}
Where
\begin{equation}
G_{\pm }(\mathbf{k}, i\omega _{n})
=\frac{1}{2} 
 \left[ \frac{1}{i\omega _{n}-\epsilon _{\mathbf{k},+}}
 \pm \frac{1}{i\omega _{n}-\epsilon _{\mathbf{k},-}}
 \right] ,
\end{equation}
%
%
\begin{equation}
 \epsilon _{\mathbf{k},\pm }
 =\xi_{\mathbf{k}}\pm \alpha 
\mid \mathbf{g_k}\mid ,
\end{equation}
\begin{equation}
 \hat{\mathbf{g}}_{\mathbf{k}}=\frac{\mathbf{g}_{\mathbf{k}}}{|\mathbf{g}_{\mathbf{k}}|}
 =\frac{1}{\sqrt{k_x^2+k_y^2}}(-k_y,k_x,0)
\end{equation}
%
%
$\omega_n=(2n +1) \pi k_B T$ is the Matsubara frequency,
and
$\sigma _{0}$ is the $2 \times 2$ unit matrix.
When the electrons
$(\mathbf{k} +\frac{\mathbf{q}}{2}, s_1)$ 
and $(-\mathbf{k}+\frac{\mathbf{q}}{2},s_2)$ make a pair,
the linearized gap equation for FFLO state is written in a $2 \times 2$ matrix form\cite{Frigeri} as
\begin{align}
 \Delta_{ss'}(\mathbf{k} +\frac{\mathbf{q} }{2}) 
 &= -k_{B}T_{c}\sum_{\mathbf{k}' ,n}\sum_{s_1,s_2}V_{\mathbf{kk} '}
 G_{ss_1}^{0}(\mathbf{k}' +\frac{\mathbf{q}}{2},i\omega _{n})  \nonumber \\
 &\times 
 \Delta_{s_1s_2}(\mathbf{k}' +\frac{\mathbf{q}}{2})G_{s's_2}^{0}(-\mathbf{k}' 
 +\frac{\mathbf{q}}{2},-i\omega _{n}).
\end{align}
The gap function is decomposed into a spin singlet part
[$\psi (\mathbf{k}+\frac{\mathbf{q}}{2})$] 
and a spin triplet 
part [$\mathbf{d}(\mathbf{k}+\frac{\mathbf{q}}{2})$],   
as
\begin{equation}
\Delta (\mathbf{k} +\frac{\mathbf{q}}{2})
=[\psi (\mathbf{k}+\frac{\mathbf{q}}{2})\sigma_{0}+\mathbf{d}(\mathbf{k}+\frac{\mathbf{q}}{2})
\cdot {\vec{\sigma }}]i{\sigma }_y .
\end{equation}
%
%
Spin triplet part and spin singlet part are mixed in general.
\begin{align}
 \mathbf{d}(\mathbf{k}+\frac{\mathbf{q}}{2})
 &=
 -k_{B}T_{c}\sum_{n,\mathbf{k}'}V_{\mathbf{kk}'} 
 \nonumber \\ 
 &\Big( [G_{-}G_{+}\hat{\mathbf{g}}-G_{+}G_{-}\hat{\mathbf{g}}']\Psi (\mathbf{k}'+\frac{\mathbf{q}}{2}) 
 \nonumber \\
 &+ G_{+}G_{+}\mathbf{d}(\mathbf{k}'+\frac{\mathbf{q}}{2}) 
 \nonumber \\
 &-iG_{-}G_{-}(\hat{\mathbf{g}}\times \hat{\mathbf{g}'})\Psi (\mathbf{k}'+\frac{\mathbf{q}}{2})
 \nonumber \\
 &+ i[G_{+}G_{-}\hat{\mathbf{g}'} + G_{-}G_{+}\hat{\mathbf{g}}]
 \times \mathbf{d}(\mathbf{k}'+\frac{\mathbf{q}}{2}) \nonumber \\ 
 &+G_{-}G_{-}[\hat{\mathbf{g}}\times {\mathbf{d}}(\mathbf{k}'+\frac{\mathbf{q}}{2})\times \hat{\mathbf{g}'} 
 -\hat{\mathbf{g}}\cdot \mathbf{d}(\mathbf{k}'+\frac{\mathbf{q}}{2})
 ]
 \Big) ,
\label{eq7}
\end{align}
where we have used the short notation for the products:
$G_{a}G_{b}=G_{a}(\mathbf{k}+\frac{\mathbf{q}}{2},
i\omega_{n})G_{b}(-\mathbf{k}+\frac{\mathbf{q}}{2}, -i\omega_{n})$
with $a,b =\pm $ and set $\mathbf{g}=\mathbf{g}_{\mathbf{k}'+\frac{\mathbf{q}}{2}}$, 
$\mathbf{g}'=\mathbf{g}_{-\mathbf{k}'+\frac{\mathbf{q}}{2}}$.
We assume that 
the density of states is constant near the Fermi surface, i.e., the density of states is $N(0)$
even if the Fermi surface is shifted by the spin-orbit coupling.
In this assumption, the particle-hole symmetry is satisfied and
the singlet and the triplet order parameters do not mix.
In other words, $\psi (\mathbf{k}'+\frac{\mathbf{q}}{2})$ terms in 
the right hand side in eq.~(\ref{eq7}) vanish.
Then we get the gap equation for triplet superconductivity as
\begin{align}
 &  \mathbf{d}(\mathbf{k}+\frac{\mathbf{q}}{2})
 \nonumber \\
 &= -k_{B}T_{c}\sum_{n,\mathbf{k}'}V_{\mathbf{kk}'} 
 \Big(G_{+}G_{+} \mathbf{d}(\mathbf{k}' + \frac{\mathbf{q}}{2}) \nonumber \\
 &+ i[G_{+}G_{-}\hat{\mathbf{g}}' 
 +G_{-}G_{+}\hat{\mathbf{g}}] 
 \times \mathbf{d}(\mathbf{k}'+\frac{\mathbf{q}}{2}) 
 \nonumber \\
 &+ G_{-}G_{-} \Big[ \hat{\mathbf{g}}\times \mathbf{d}(\mathbf{k}'+\frac{\mathbf{q}}{2})
 \times \hat{\mathbf{g}}'-
 \hat{\mathbf{g}}\cdot \mathbf{d}(\mathbf{k}'+\frac{\mathbf{q}}{2})
 \hat{\mathbf{g}}' \Big]
 \Big).
\end{align}
In the following 
we set $\mathbf{q}=(0,0,k_{F}q)$ and 
\begin{equation}
 \xi _{\pm \mathbf{k}+\frac{\mathbf{q}}{2}} 
= \tilde{\xi}_{\mathbf{k}}\pm \frac{\hbar ^{2}k_{F}^{2}q \cos\theta }{2m},
\end{equation}
where 
$\tilde{\xi}_{\mathbf{k}} = \xi_{\mathbf{k}}+ \frac{\hbar^2 k_F^2 q^2}{8m}$.
Then we find the transition temperature for spin triplet pairing is given by
\begin{align}
 &\ln  (\frac{T_c}{T_{c0}}) 
 =-\pi k_{B}T_{c} 
 \langle {\rm Im}\sum_{n=0}^{n_{c}-1}
 \nonumber \\
 &\Big( 
 (f_{1-}+f_{2-}+f_{2+}+f_{1+} -\frac{2}{i \left| \omega _{n} \right|}) 
 {\left| {\mathbf{d}(\mathbf{k}+\frac{\mathbf{q}}{2}) }\right| }^2 
 \nonumber \\
 &- (f_{1-}-f_{2-}-f_{2+}+f_{1+})
 \nonumber \\ 
 &\times \Big[ 2(
 \hat{\mathbf{g}'}
 \cdot \mathbf{d} (\mathbf{k}+\frac{\mathbf{q}}{2}))
 (\hat{\mathbf{g}}
 \cdot \mathbf{d}(\mathbf{k}+\frac{\mathbf{q}}{2}))  
 \nonumber \\
 & \ -\hat{\mathbf{g}'}
 \cdot \hat{\mathbf{g}}
 \left| \mathbf{d}(\mathbf{k}+\frac{\mathbf{q}}{2})\right| ^2 \Big]
 \Big)
 \rangle_{\mathbf{k}},
\label{eq10}
\end{align}
where 
\begin{align}
 f_{1\pm }
 &=\frac{1}{\frac{\hbar ^{2}k_{F}^{2}q \cos \theta }{m}+2i\mid \omega _{n}\mid
 \pm \alpha (\mid {\mathbf{g}}\mid 
 -\mid {\mathbf{g}'}\mid )}, 
 \\
 f_{2\pm }
 &=\frac{1}{\frac{\hbar ^{2}k_{F}^{2}q \cos \theta }{m}+2i\mid \omega _{n}\mid 
\pm \alpha (\mid {\mathbf{g}}\mid 
+\mid {\mathbf{g}'}\mid )},
\end{align}
and $T_{c0}$ is the transition temperature for $\alpha=0$, i.e.,
$k_{B}T_{c0}=\epsilon _{c}$exp$(-1/\lambda _{t})$ with 
\begin{equation}
 \lambda _{t}\mathbf{d}(\mathbf{k})
 =-N(0)\big< V_{\mathbf{kk}'}\mathbf{d}(\mathbf{k}')\big>_{\mathbf{k}'}.
\end{equation}



In the following we assume  $\mathbf{q} \parallel \hat{z}$, for simplicity.
Then $\mathbf{g}_{\mathbf{k}+\frac{\mathbf{q}}{2}}$   
is independent of $\mathbf{q}$,
$\mathbf{g}_{\mathbf{k}+\frac{\mathbf{q}}{2}}=\sqrt{\frac{3}{2}}\frac{1}{k_F}(-k_{y},k_{x},0)$. 
In the next section, we calculate the transition temperature as a function of $\alpha $ 
in $\mathbf{q}=0$ for
the order parameters shown in Table \ref{table1} \, which are the basis functions 
for the $C_{4v}$ point group suggested by A. Sergienko and S. H. Curnoe\cite{Sergienko} and 
P. A. Frigeri et al.\cite{Frigeri}. 
%
%
\begin{table}
\begin{center}
\begin{tabular}{cl}
 \hline
 Basis function  & \hspace{0.5cm}       Order parameter \\ 
 \hline
 $A_1$ & \hspace{0.5cm} $\mathbf{d}(\mathbf{k})=\sqrt{\frac{3}{2}}\frac{1}{k_F}(k_x, k_y, 0)$ \\
       & \hspace{0.5cm} $\mathbf{d}(\mathbf{k})=                  \frac{1}{k_F}(k_x, k_y, k_z)$ \\
       & \hspace{0.5cm} $\mathbf{d}(\mathbf{k})=\sqrt{\frac{7}{2}}\frac{1}{k_F^3}(k_x^3, k_y^3, 0)$ \\
 $A_2$ & \hspace{0.5cm} $\mathbf{d}(\mathbf{k})=\sqrt{\frac{3}{2}}\frac{1}{k_F}(-k_y, k_x, 0)$ \\
 $B_1$ & \hspace{0.5cm} $\mathbf{d}(\mathbf{k})=\sqrt{\frac{3}{2}}\frac{1}{k_F}(k_x, -k_y, 0)$ \\
 $B_2$ & \hspace{0.5cm} $\mathbf{d}(\mathbf{k})=\sqrt{\frac{3}{2}}\frac{1}{k_F}(k_y, k_x, 0)$ \\  
 $E$   & \hspace{0.5cm} $\mathbf{d}(\mathbf{k})=\sqrt{3}          \frac{1}{k_F}(k_z, 0, 0)$ \\
       & \hspace{0.5cm} $\mathbf{d}(\mathbf{k})=\sqrt{3}          \frac{1}{k_F}(0, k_z, 0)$ \\ \hline
\end{tabular}\end{center}
\caption
{Order parameters in $C_{4v}$ point group. The order parameters are normalized as
$\langle |\mathbf{d}(\mathbf{k}) |^2 \rangle =1$ for the spherical Fermi surface.}
\label{table1} 
\end{table}
\begin{figure}[tbh]
\includegraphics[width=0.5\textwidth]{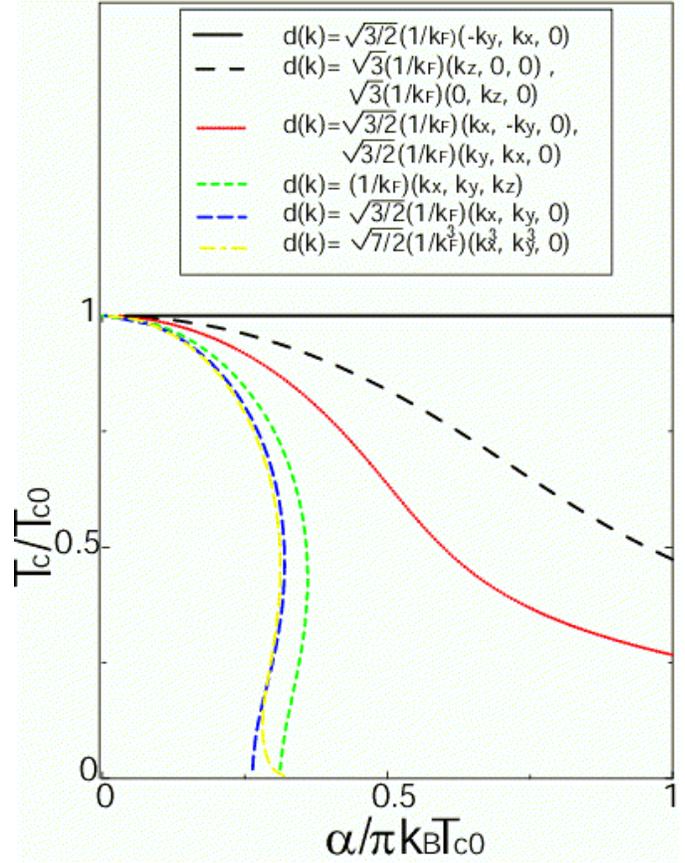}
\caption
{ Transition temperature  as a function of the strength of the
spin-orbit coupling $\alpha $ 
for some types of spin triplet superconductivity with $\bar{q}=0$
and
$\mathbf{g}_{\mathbf{k}}=\sqrt{\frac{3}{2}}\frac{1}{k_F}(-k_{y}, k_{x}, 0)$.
When $\mathbf{d}(\mathbf{k})\parallel \mathbf{g_k}$, $T_c$ is independent of $\alpha$.
For the states with $\mathbf{d}(\mathbf{k}) \propto (k_z,0,0), (0,k_z,0),
(k_x,-k_y,0)$ and $(k_y,k_x,0)$, $T_c$  is reduced by $\alpha$
as $T_c \propto 1/\alpha$. For $\mathbf{d}(\mathbf{k}) \propto (k_x,k_y,k_z)$ and $(k_x,k_y,0)$
$T_{c}$ becomes zero at the critical value of $\alpha$. For $\mathbf{d}(\mathbf{k}) \propto
(k_x^3, k_y^3,0)$, $T_c$ is reduced sharply near $\alpha \approx 0.3 \pi k_B T_{c0}$ and
$T_c \propto \alpha^{-19}$ at large $\alpha$. }
\label{fig.1}
\end{figure}
\vspace{0.5cm}

\section{Transition temperature of the uniform state}
Fig. 1 display the transition temperature as a function of $\alpha $ for some
types of spin triplet superconductivity with $\bar{q}=0$.
The transition temperature of the 
state with $\mathbf{d}(\mathbf{k}) \parallel \mathbf{g}_{\mathbf{k}}$
does not depend on $\alpha$.
As shown in Appendix,
we obtain that the transition temperature for 
 $\mathbf{d}(\mathbf{k}) =\sqrt{3}\frac{1}{k_F} (k_z,0,0)$ and  
 $\mathbf{d}(\mathbf{k}) =\sqrt{3}\frac{1}{k_F} (0,k_z,0)$\ 
(the two-dimensional representation $E$)  
approaches to zero as proportional to
$\frac{\pi k_{B}T_{c0}}{\alpha}$, for large $\alpha$ .
The transition temperatures for  
 $\mathbf{d}(\mathbf{k})= \sqrt{\frac{3}{2}} \frac{1}{k_F}(k_x, -k_y, 0)$
and
 $\mathbf{d}(\mathbf{k})= \sqrt{\frac{3}{2}} \frac{1}{k_F} (k_y, k_x, 0)$
 also go to zero as proportional to
$\frac{\pi k_{B}T_{c0}}{\alpha}$, for large $\alpha$.
The transition temperature for 
 $\mathbf{d}(\mathbf{k}) = \frac{1}{k_F} (k_x, k_y, k_z)$ 
is zero at $\alpha \approx 0.31 \pi k_B T_{c0}$
and  the transition temperature for 
 $\mathbf{d}(\mathbf{k}) = \sqrt{\frac{3}{2}} \frac{1}{k_F}(k_x, k_y, 0)$
is zero at  $\alpha \approx 0.26 \pi k_B T_{c0}$.
Interestingly, the transition temperature for  
 $\mathbf{d}(\mathbf{k}) = \sqrt{\frac{7}{2}}  \frac{1}{k_F}(k_x^3, k_y^3, 0)$
depends on $\alpha$ in the similar way as the transition temperature for
  $\mathbf{d}(\mathbf{k}) = \sqrt{\frac{3}{2}} \frac{1}{k_F}(k_x, k_y, 0)$,
but it 
approaches to zero as  $(\frac{\alpha}{\pi k_{B}T_{c0}})^{-19}$ for large $\alpha$.
Curves of $T_c$ vs $\alpha$ for 
$\mathbf{d}(\mathbf{k})= \frac{1}{k_F}(k_x, k_y, k_z)$, 
$\mathbf{d}(\mathbf{k})=\sqrt{\frac{3}{2}} \frac{1}{k_F}(k_x, k_y,0)$ 
and 
$\mathbf{d}(\mathbf{k})=\sqrt{\frac{7}{2}} \frac{1}{k_F}(k_x^3, k_y^3,0)$
are reentrant and they
are reminiscent to the transition temperature for the singlet superconductivity
as a function of  external magnetic field, where 
only the Pauli pair breaking effects are taken into account and the 
orbital effects are neglected\cite{Saint-James,Takada}.
%
%
\begin{figure}[tbh]
\begin{center}
\includegraphics[width=0.45\textwidth]{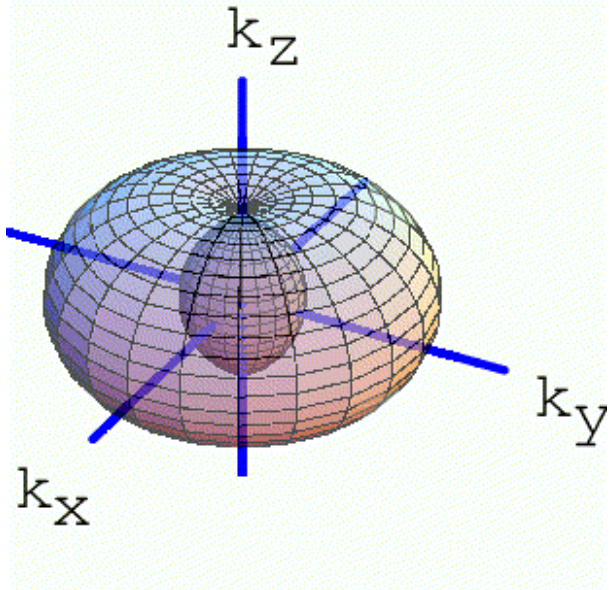}
\end{center}
\caption
{Fermi surfaces in the presence of the spin-orbit coupling.
Spherical Fermi surface ($\xi_{\mathbf{k}}=0$) is split into two surfaces, $\epsilon _{+}=0$ and
$\epsilon _{-}=0$. 
}
\label{fig2}
\end{figure}
\begin{figure}[tbh]
\includegraphics[width=0.45\textwidth]{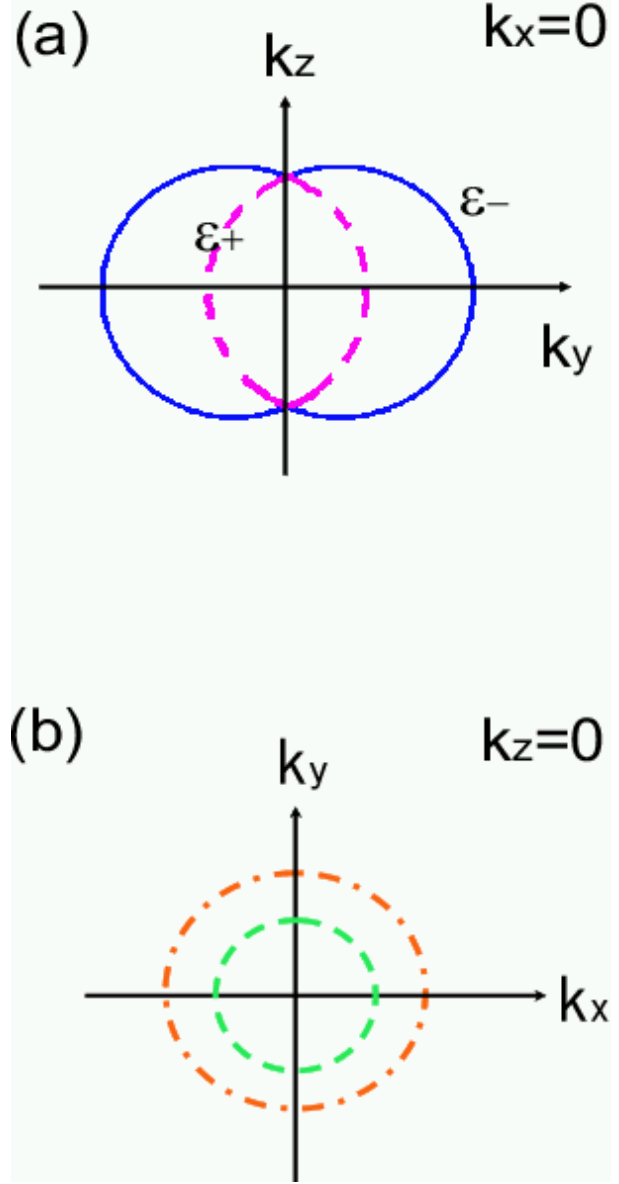}
\caption
{Cross section of the Fermi surface with the plane  $k_x=0$ (a) and the plane $k_z=0$ (b). 
}
\label{fig3}
\end{figure}
%


%

Here we summarize the known results for the Pauli pair breaking effect in
 the spin singlet superconductivity\cite{Saint-James,Takada}.
In the case of Pauli pair breaking for the spin singlet superconductivity
 Fermi surfaces is split into two concentric spheres  by the external magnetic field.
If we assume the second order phase transition into the superconducting state
with $\mathbf{q}=0$,
the reentrance is obtained. However, this reentrance cannot be realized, since the first order transition 
to the superconductivity with $\mathbf{q}=0$ is expected in the region where
the reentrance might exist. 
The first order transition  into the $\mathbf{q}=0$ 
superconductivity from the normal state does not realized neither, since
it is covered by the second order transition into the pairing with finite  $\mathbf{q}$ 
(FFLO state).

The similar situation is expected in the case of the pair breaking due to an
anisotropic spin-orbit coupling.
In case of spin-orbit coupling 
electron energies for the Hamiltonian (eq.~(\ref{H0})) are   
\begin{equation}
 \epsilon_{\pm } = \xi_{\mathbf{k}}\pm \alpha \sqrt{k_x^2+k_y^2},
\end{equation}  
and the  eigenstates are
\begin{equation}
 \varphi _{\pm } = \frac{1}{\sqrt{2}}\left(
 \begin{array}{@{\,}cccc@{\,}}
 1 \\
 \mp \frac{k_{y}-ik_{x}}{\sqrt{k_{y}^2+k_{x}^{2}}}
 \end{array}
 \right),
\end{equation}
where the upper and the lower components are the up and down spins.
The Fermi surface is split as shown in Fig.~\ref{fig3}.
In the present case  Fermi surfaces is split due to spin-orbit coupling 
as two spheres with different centers as
in Fig. 2 and Fig. 3.
The first order transition into the spin triplet state with $\mathbf{q}=0$ might occur, but 
we do not consider that possibility in this paper.
The second order transition into the state with $\mathbf{q} \neq 0$ in the next section. 
%
%

\section{Nonuniform state of the spin triplet pairing due to anisotropic spin-orbit coupling} 

%
%
%
%
%
\begin{figure}[tbh]
\begin{center}
\includegraphics[width=0.45\textwidth]{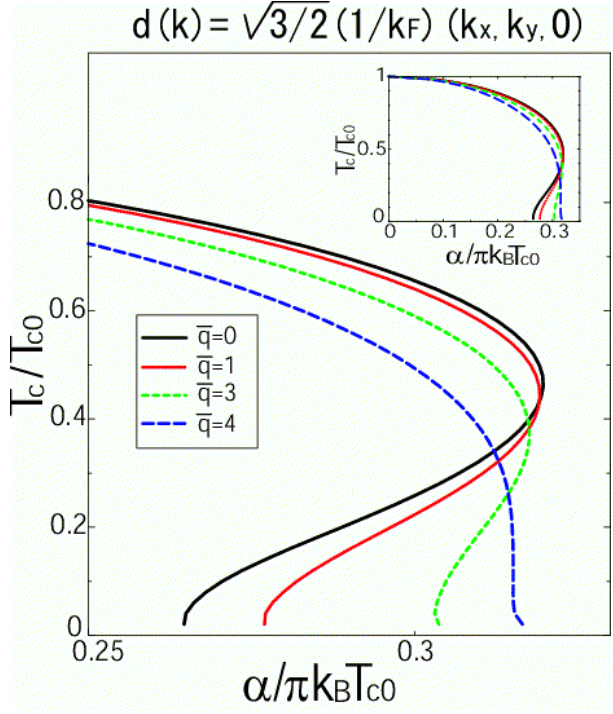}
\end{center}
\caption
{ Transition temperature  as a function of $\alpha $ 
for triplet superconductivity with 
$\mathbf{g}_{\mathbf{k}}=\sqrt{\frac{3}{2}}\frac{1}{k_F}(-k_{y}, k_{x}, 0)$
and 
$\mathbf{d}(\mathbf{k})=\sqrt{\frac{3}{2}}\frac{1}{k_F}(k_x, k_y, 0)$.
}
\label{fig4}
\end{figure}

%
%
In  Fig.~\ref{fig4}, we plot transition temperatures of the states 
$\mathbf{d}(\mathbf{k})=\sqrt{\frac{3}{2}}\frac{1}{k_F}(k_x,k_y,0)$,
as a function of $\alpha $ for some  values of $\bar{q}$, 
which is defined  by
\begin{equation}
 \bar{q}=\frac{\hbar ^{2}k_{F}^{2}}{\alpha m}q.
\end{equation}
For $\bar{q}=0$, the transition temperature shows a reentrance near 
$\alpha/(\pi k_B T_{c0}) \approx 0.3$
($\alpha_c \leq \alpha \leq \alpha_0$,
$\alpha_c \approx 0.263 \pi k_B T_{c0}$,   and $\alpha_0 \approx 0.320\pi k_B T_{c0}$).
%
%
%
%
%
%
%
The reentrance never occurs, 
since the transition curve is calculated by assuming the second order transition. 
When the temperature is lowered at fixed $\alpha$ ($\alpha_c \leq \alpha \leq \alpha_0$), 
the order parameter is finite below the higher transition temperature 
and the lower transition temperature is spurious, as in the case of the 
Pauli pair breaking in the spin singlet superconductivity.
For the finite value of $\bar{q}$ the critical value $\alpha_c$ at $T_{c0}=0$ becomes larger. 
However, the critical value of $\alpha$ never exceed $\alpha_0\approx 0.32 \pi k_B T_{c0}$ for
$\bar{q}=0$.
Therefore, contrary to the Pauli pair breaking in the spin singlet pairing,
the spin orbit coupling will not stabilize
the nonuniform state of $\mathbf{d}(\mathbf{k})=\sqrt{\frac{3}{2}}\frac{1}{k_F} (k_x, k_y, 0)$.
%
%
\begin{figure}[tbh]
\includegraphics[width=0.45\textwidth]{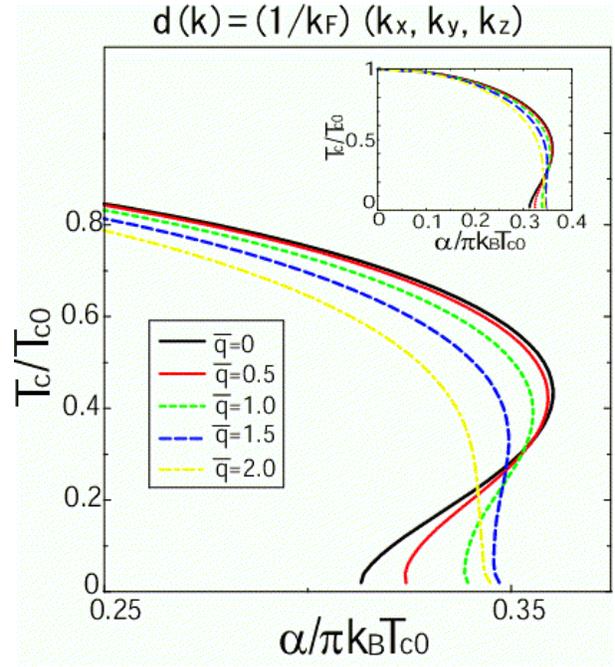}
\caption{ Transition temperature  as a function of $\alpha $ 
for triplet superconductivity with 
$\mathbf{g}_{\mathbf{k}}=\sqrt{\frac{3}{2}}\frac{1}{k_F}(-k_{y}, k_{x}, 0)$
and 
$\mathbf{d}(\mathbf{k})=\frac{1}{k_F}(k_x, k_y, k_z)$.
}
\label{fig5}
\end{figure}

\begin{figure}[tbh]
\includegraphics[width=0.45\textwidth]{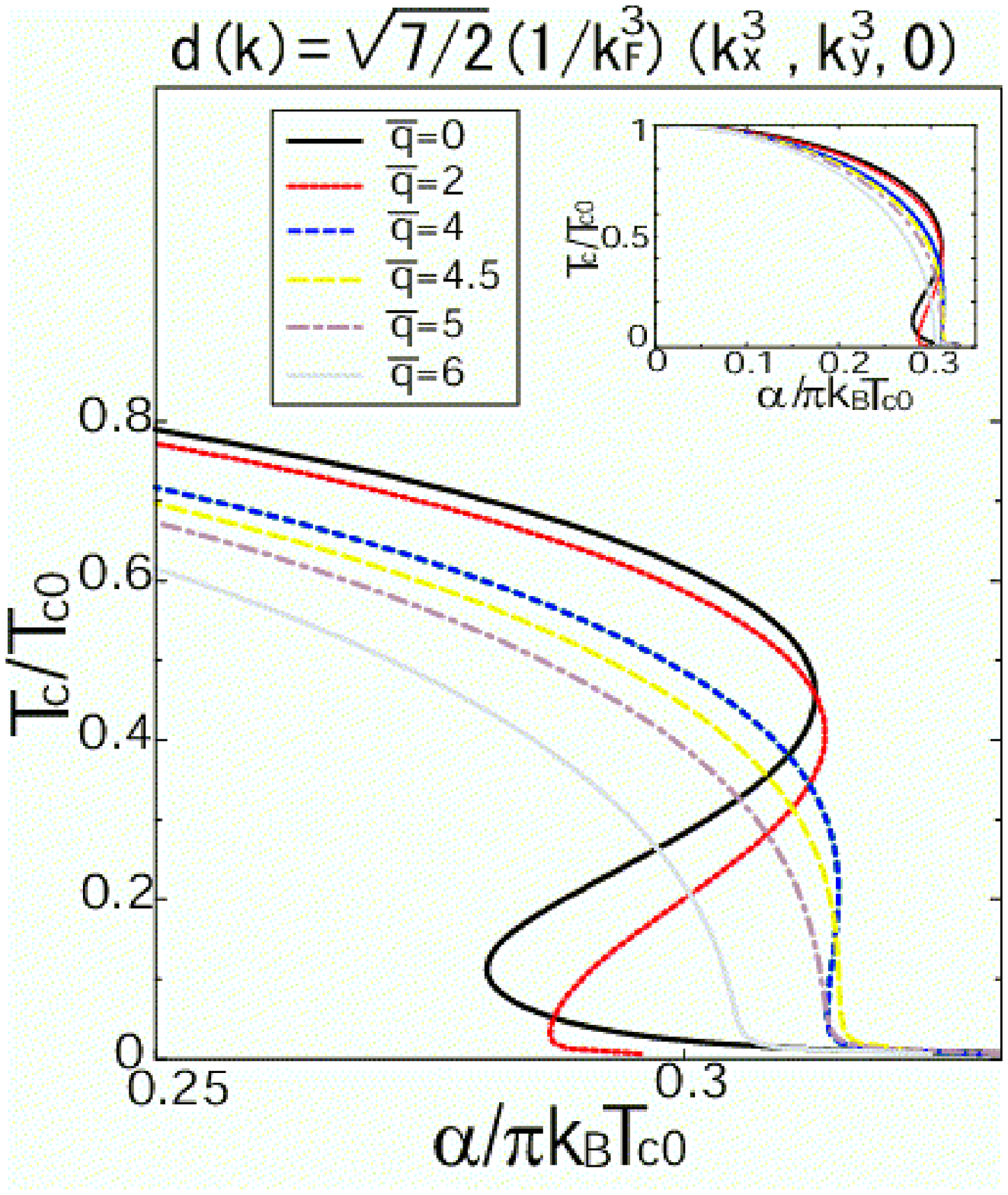}
\caption{ Transition temperature  as a function of $\alpha $ 
for triplet superconductivity with 
$\mathbf{g}_{\mathbf{k}}=\sqrt{\frac{3}{2}}\frac{1}{k_F}(-k_{y}, k_{x}, 0)$
and
$\mathbf{d}(\mathbf{k})=\sqrt{\frac{7}{2}}\frac{1}{k_F^3}(k_x^3, k_y^3, 0)$.
}
\label{fig6}
\end{figure}

We obtain the similar $\alpha$ dependence of $T_c$ for
the superconducting state  of $\mathbf{d}(\mathbf{k})=\frac{1}{k_F} (k_x, k_y, k_z)$
as shown in Fig.~\ref{fig5}. 
The reentrance seen in the region $0.263 \lesssim \frac{\alpha}{\pi k_B T_{c0}} \lesssim 0.360$ 
when $\bar{q}=0$.
When $\bar{q} \neq 0$, the reentrant region becomes small, 
but the superconducting state is not 
realized in the region $\alpha \gtrsim 0.360$. Then the nonuniform superconducting state is not stabilized 
for this order parameter. 

For the superconducting state with 
$\mathbf{d}(\mathbf{k})=\sqrt{\frac{7}{2}} \frac{1}{k_F^3} (k_x^3, k_y^3,0)$,
we find that the nonuniform state is stabilized in the region 
$\frac{\alpha}{\pi k_B T_{c0}} \gtrsim 0.312$ as shown in Fig.~\ref{fig6}. 
In this type of the superconductivity $T_c$ is reentrant when $\alpha \lesssim 0.312$
and $T_c \propto \alpha^{-19}$ for $T_c \ll T_{c0}$ and $\alpha \gg \pi k_B T_{c0}$.
As seen in Fig.~\ref{fig6}, the nonuniform state with $0 < \bar{q} \lesssim 4.5$ is stabilized 
in the when $\alpha \gtrsim 0.312$.

%
\vspace{0.5cm}
\section{Conclusion}
We study the pair breaking effect due to the spin-orbit coupling for the spin triplet superconductivity.
We have studied various type of spin triplet superconductivity, which is characterized by the $\mathbf{d}$
vector,
the states with $\mathbf{d}(\mathbf{k}) = \sqrt{\frac{3}{2}} \frac{1}{k_F} (k_x, k_y,0)$,
$ \frac{1}{k_F} (k_x, k_y, k_z)$, and $\sqrt{\frac{7}{2}} \frac{1}{k_F^3} (k_x^3, k_y^3,0)$
show the reentrant curves of $T_c$ as a function of $\alpha$.
Although the non-uniform state ($\bar{q}\neq 0$, FFLO state) is not stabilized in the former two states,
we find that the nonuniform state is stabilized in the latter state.
The region of $\alpha$ where the nonuniform state is stabilized is rather small. 
In this study we have assumed the spherical Fermi surface. 
If we consider the Fermi surface 
with different shape with lower dimensionality (quasi-two-dimension or quasi-one dimension)
 or better nesting condition of the Fermi surface, FFLO state is expected to appear
in wider region of parameters, as is the case 
in the ordinary FFLO state caused by the Pauli pair breaking\cite{Shimahara}.%

If the FFLO state is realized in the absence of magnetic field, the amplitude of the energy gap
varies in space and the specific heat and other properties will
depend on temperature as a power low.
The experimentally observed properties of CePt$_3$Si, i.e. spin triplet superconductivity with 
power-low temperature dependences of penetration depth and thermal conductivity, are consistent with the
nonuniform spin triplet state due to the spin-orbit coupling.  

%
%
\section*{Acknowledgement}
We thank H. Shimahara and N. Hayashi for useful discussions.
H.T and H.K. thank M. Sigrist, D. Agterberg and Y. Yanase for fruitful comments.


\appendix
\section{Transition temperature in the case of strong pair breaking due to spin-orbit coupling}
In this appendix
we calculate the transition temperature of the uniform 
triplet superconductivity in eq.~(\ref{eq10}) 
for $\mathbf{q}=0$ and $\frac{T_c}{T_{c0}}\rightarrow 0$.
For $\mathbf{q}=0$, eq.~(\ref{eq10}) becomes 
\begin{align}
 \ln(\frac{T_{c}}{T_{c0}})
 &= -2\pi k_{B}T_{c}
 \langle [\mid \mathbf{d}(\mathbf{k})\mid ^{2}-
 (\hat{\mathbf{g}}_{\mathbf{k}}\cdot \mathbf{d}(\mathbf{k}))^{2}] \nonumber \\
 &\times
  \mathrm{Im} \sum_{n=0}^{n_{c}-1}
 \Big(-\frac{1}{i\mid {\omega _n}\mid}
  +\frac{1}{2i\mid {\omega _n}\mid
  +2\alpha \mid \mathbf{g_k}\mid }  
 \nonumber \\
 & +\frac{1}{2i\mid {\omega _n}\mid
 -2\alpha \mid \mathbf{g_k}\mid } 
 \Big)\rangle _{\mathbf{k}}.
\end{align}
We set $\rho _{\mathbf{k}}=\frac{\alpha \mid \mathbf{g_k}\mid}{\pi k_{B}T_{c0}}$. 
As we assume that Fermi surface is spherical,
\begin{align}
 k_x &= k_F \sin \theta \cos\phi \\
 k_y &= k_F \sin \theta \sin\phi \\
 k_z &= k_F \cos \theta
\end{align}
and the average over the Fermi surface is performed as
\begin{equation}
  \langle \cdots  \rangle_{\mathbf{k}} 
=
\frac{1}{4\pi }\int _{0}^{2\pi }d\phi 
\int _{0}^{\pi }d\theta {\sin}\theta \cdots
\end{equation}
%
Then the transition temperature is obtained by
\begin{align}
 \ln (\frac{T_c}{T_{c0}})
 &=
 -\frac{1}{2\pi }
 \int _{0}^{2\pi }d\phi \int _{-1}^{1}d(\cos \theta)
 [\mid \mathbf{d}(\mathbf{k})\mid ^2-(\hat{\mathbf{g}}_{\mathbf{k}}\cdot 
 \mathbf{d}(\mathbf{k}))^2] \nonumber \\
 &\times
 \frac{1}{2}{\rm Re} 
 \Big[ \Psi (n_c+\frac{1}{2})-\Psi (\frac{1}{2})
 \nonumber \\
 & -\Psi (n_c+\frac{1}{2}+i\frac{\rho_{\mathbf{k}}}{2})
   +\Psi (\frac{1}{2}+i\frac{\rho_{\mathbf{k}}}{2}) \Big],
\label{eqA2}
\end{align}
where $\Psi(n_c+\frac{1}{2})$ is the digamma function.
Since $n_c \gg 1$, we set
$\Psi (n_{c}+\frac{1}{2})-\Psi (n_c + \frac{1}{2} + i \frac{\rho_{\mathbf{k}}}{2}) \approx 0$
in eq.~(\ref{eqA2}).
%
%
At low temperatures ($T_c \ll T_{c0}$),
$\rho_{\mathbf{k}} \gg 1$. Then  
\begin{align} 
 \mathrm{Re}\left( \Psi (\frac{1}{2}+i\frac{\rho_{\mathbf{k}}}{2}) \right)
 &\approx \mathrm{Re}\left( \Psi (i\frac{\rho_{\mathbf{k}}}{2}) \right) \nonumber \\
 &\approx \mathrm{Re}\left( {\ln}(i\frac{\rho_{\mathbf{k}}}{2}) \right) \nonumber \\
 &= \ln(\frac{\rho_{\mathbf{k}}}{2} ) \nonumber \\
 &= \ln(\frac{\alpha \mid \mathbf{g_k}\mid }{2\pi k_{B}T_{c}})
\end{align}
and 
the transition temperature in the case of $T_c \ll T_{c0}$ is obtained by
\begin{align} 
 {\ln}(\frac{T_c}{T_{c0}})
 &\approx -\frac{1}{4\pi }
 \int _{0}^{2\pi }d\phi \int _{-1}^{1}d(\cos \theta)
 \left[ \mid \mathbf{d}(\mathbf{k})\mid ^2-(\hat{\mathbf{g}}_{\mathbf{k}}\cdot 
 \mathbf{d}(\mathbf{k}))^2 \right]
 \nonumber \\ 
 &\times \ln \big( 2e^{\gamma }\frac{\alpha \mid \mathbf{g_k}\mid}{\pi k_{B}T_{c}}\big),
 \label{eq16}
\end{align}
where we have used
$\Psi (\frac{1}{2})=-{\ln}(4e^{\gamma })$ and 
$\gamma $ is the Euler constant.

\subsection{$\mathbf{d}(\mathbf{k})=\sqrt{3}\frac{1}{k_F}(k_{z}, 0, 0)$}
 First, we study the order parameter 
$\mathbf{d}(\mathbf{k})=\sqrt{3}\frac{1}{k_F}(k_{z}, 0, 0)$.
In this case
\begin{equation}
\mid \mathbf{d}(\mathbf{k})\mid ^2-(\hat{\mathbf{g}}_{\mathbf{k}}\cdot \mathbf{d}(\mathbf{k}))^2
=\frac{3}{2}\cos^2 \theta (1+{\cos}2\phi ).
 \label{eq17}
\end{equation}
Substituting eq.~(\ref{eq17}) into eq.~(\ref{eq16}), we obtain  
\begin{align}
 {\ln}(\frac{T_c}{T_{c0}})
 &= -\frac{1}{4\pi }
 \int _{0}^{2\pi }d\phi \int _{-1}^{1}dz
 \frac{3}{2}z^2 (1+{\cos}2\phi )
 \nonumber \\
 & \ \times \ln \big( \frac{\sqrt{6(1-z^2)}e^{\gamma }\alpha }{\pi k_{B}T_{c}}\big) \nonumber \\
 &=-\frac{3}{2}\int_{0}^{1}dz
 \left[ z^2{\ln}(\frac{\sqrt{6}e^{\gamma }\alpha }
 {\pi k_{B}T_c})+\frac{1}{2}z^2{\ln}(1-z^2)
 \right] .
\end{align}
We perform the integration over  $k_z$  using 
\begin{equation}
 \int _{0}^{1}dz z^{2}{\ln}(1-z^2)=\frac{2}{9}({\ln}8-4),
\end{equation}
and  we obtain
\begin{equation}
\ln (\frac{T_c}{T_{c0}})
 =
 -\frac{1}{2}\left[ \ln (\frac{\alpha }{\pi k_{B}T_c})
 +\ln (2\sqrt{6}e^{\gamma -\frac{4}{3}}) \right] .
\end{equation}  
Finally we obtain that the transition temperature 
for $T_c \ll T_{c0}$ is obtained as
\begin{align}
 \frac{T_c}{T_{c0}}
 &=\frac{1}{2\sqrt{6}e^{\gamma -\frac{4}{3}}} \frac{\pi k_{B}T_{c0}}{\alpha }\\
 &\approx 0.435  \frac{\pi k_{B}T_{c0}}{\alpha },
\end{align}
i.e. $T_c$ is reduced as $\alpha^{-1}$ in the strong spin-orbit coupling case
 ($\alpha \gg 1$).

%
\subsection{$\mathbf{d}(\mathbf{k})=\sqrt{\frac{3}{2}}\frac{1}{k_F}(k_x, -k_y, 0)$}
In this subsection we study  the case 
of $\mathbf{d}(\mathbf{k})=\sqrt{\frac{3}{2}}\frac{1}{k_F}(k_x, -k_y, 0)$.
In this case we obtain
\begin{equation}
\mid \mathbf{d}(\mathbf{k})\mid ^2-(\hat{\mathbf{g}}_{\mathbf{k}}\cdot \mathbf{d}(\mathbf{k}))^2
=\frac{3}{4}(1-\cos^2\theta)(1+{\cos}4\phi ),
\label{eq23}
\end{equation}
and we perform the  integration over  $\phi $ and $k_z$ in  eq.~(\ref{eq16}).
We use 
\begin{equation}
\int _{0}^{1} dz {\ln}(1-z^2)=-2+{\ln}4
\end{equation}
 and
\begin{equation}
 \int _{0}^{1} dz z^{2}{\ln}(1-z^2)=\frac{9}{2}({\ln}8-4).
\end{equation}
In this case we obtain,
\begin{align}
 {\ln}(\frac{T_{c}}{T_{c0}})
 &=-\frac{1}{4\pi }\cdot \frac{3}{4}\int_{0}^{2\pi }d\phi 
 \int _{-1}^{1} dz (1-z^2)
 \nonumber \\
 &\times (1+{\cos}4\phi ) 
 {\ln}(\frac{\sqrt{6(1-z^2)}e^{\gamma}
 \alpha }{\pi k_{B}T_c}) \nonumber \\
 &=-\frac{1}{2}{\ln}(\frac{\sqrt{6}e^{\gamma }\alpha }{\pi k_{B}T_c})
 -\frac{3}{8}(-2+{\ln}4) \nonumber \\
 & +\frac{3}{8} \cdot \frac{2}{9}(3{\ln}2-{\ln}e^4).
\end{align}
This equation is rewritten as
\begin{equation}
 {\ln}\left( (\frac{T_c}{T_{c0}})^2(\frac{\alpha }{\pi k_{B}T_{c}}) \right)
 =
 {\ln}\big( \frac{e^{\frac{5}{6}-\gamma }}{2\sqrt{6}}\big). 
\end{equation}
Finally we obtain 
\begin{align}
 \frac{T_c}{T_{c0}}
 &= \frac{1}{2\sqrt{6}e^{\gamma -\frac{5}{6}}}
 \frac{\pi k_{B}T_{c0}}{\alpha} \nonumber \\
 &\approx 0.263\frac{\pi k_{B}T_{c0}}{\alpha} .
\end{align}

\subsection{$\mathbf{d}(\mathbf{k})=\sqrt{\frac{3}{2}}\frac{1}{k_F}(k_y, k_x, 0)$}
For
 $\mathbf{d}(\mathbf{k})=\sqrt{\frac{3}{2}}\frac{1}{k_F}(k_y, k_x, 0)$
we obtain
\begin{equation}
\mid \mathbf{d}(\mathbf{k})\mid ^2-(\hat{\mathbf{g}}_{\mathbf{k}}\cdot \mathbf{d}(\mathbf{k}))^2
=\frac{3}{4}(1-\cos^2\theta)(1 - {\cos}4\phi ) .
\label{eq23b}
\end{equation}
Since the term proportional to $\cos 4\phi$ vanish in the integration over $\phi$, we get the same
$\alpha$ dependence of $T_c$ as the case 
of $\mathbf{d}(\mathbf{k})=\sqrt{\frac{3}{2}}\frac{1}{k_F}(k_x, -k_y, 0)$.

\subsection{$\mathbf{d}(\mathbf{k})= \frac{1}{k_F}(k_x, k_y, k_z)$}
When $\mathbf{d}(\mathbf{k})=(k_x, k_y, k_z)$,
$\hat{\mathbf{g}}_{\mathbf{k}}\cdot \mathbf{d}(\mathbf{k})=0$
and 
\begin{equation}
 \mid \mathbf{d}(\mathbf{k})\mid ^2-(\hat{\mathbf{g}}_{\mathbf{k}}
 \cdot \mathbf{d}(\mathbf{k}))^2=1.
\label{eqA20}
\end{equation}
We substitute eq.~(\ref{eqA20})  into eq.~(\ref{eq16}), 
and obtain
\begin{align} 
 {\ln}(\frac{T_c}{T_{c0}})
 &=-\frac{1}{4\pi }\int _{0}^{2\pi }
 d\phi \int _{-1}^{1}dz
 {\ln}(\frac{\sqrt{6(1-z^2)}e^{\gamma }\alpha }{\pi k_{B}T_c}) \nonumber \\
 &=-{\ln}(\frac{\alpha }{\pi k_{B}T_c})-{\ln}(2\sqrt{6}e^{\gamma -1}) .
\label{eqA21}
\end{align}
We obtain
\begin{align}
 \frac{\alpha }{\pi k_{B}T_{c0}}
 &=\frac{1}{2\sqrt{6}e^{\gamma -1}} 
 \approx 0.31
\end{align}
from eq.~(\ref{eqA21}),
i.e. $T_c =0$ at $\alpha \approx 0.31 \pi k_B T_{c0}$.

\subsection{$\mathbf{d}(\mathbf{k})=\sqrt{\frac{3}{2}}\frac{1}{k_F}(k_x, k_y, 0)$}
When $\mathbf{d}(\mathbf{k})=\sqrt{\frac{3}{2}}\frac{1}{k_F}(k_x, k_y, 0)$,
$\hat{\mathbf{g}}_{\mathbf{k}}\cdot \mathbf{d}(\mathbf{k})=0$,
\begin{equation}
\mid \mathbf{d}(\mathbf{k})\mid ^2-(\hat{\mathbf{g}}_{\mathbf{k}}\cdot \mathbf{d}(\mathbf{k}))^2=
\frac{3}{2}(1-\cos^2\theta) .
\end{equation}
We substitute this into eq.~(\ref{eq16}), 
\begin{align}
 {\ln}(\frac{T_c}{T_{c0}})
 &=-\frac{1}{4\pi }\int _{0}^{2\pi }
 d\phi \int _{-1}^{1}dz \frac{3}{2}
 {\ln}(\frac{\sqrt{6(1-z^2)}e^{\gamma }\alpha }{\pi k_{B}T_c}) 
 \nonumber \\
 &=-{\ln}(\frac{\alpha }{\pi k_{B}T_c})-{\ln}(2\sqrt{6}e^{\gamma -\frac{5}{6}}) .
\end{align}
Then we obtain 
\begin{align}
 \frac{\alpha }{\pi k_{B}T_{c0}}
 &=\frac{1}{2\sqrt{6}e^{\gamma -\frac{5}{6}}} 
 \approx 0.26,
\end{align}
i.e. 
 $T_c =0$ at $\alpha \approx 0.26 \pi k_B T_{c0}$.

\subsection{$\mathbf{d}(\mathbf{k})=\sqrt{\frac{7}{2}}\frac{1}{k_F^3}(k_{x}^{3}, k_{y}^{3}, 0)$}
When $\mathbf{d}(\mathbf{k})=\sqrt{\frac{7}{2}}\frac{1}{k_F^3}(k_{x}^{3}, k_{y}^{3}, 0)$,
\begin{align}
 &\mid \mathbf{d}(\mathbf{k})\mid ^2-(\hat{\mathbf{g}}_{\mathbf{k}}\cdot \mathbf{d}(\mathbf{k}))^2
 \nonumber \\
 &= \frac{7}{64}(1-\cos^2\theta)^{3}
 (19+12 \cos 4\phi +\cos 8\phi ).
\end{align}
Substitute it into  eq.~(\ref{eq16}), we obtain
\begin{align} 
 {\ln}(\frac{T_{c}}{T_{c0}})
 &=-\frac{1}{4\pi }\frac{7}{64}\int_{0}^{2\pi }d\phi 
 \int _{-1}^{1} dz (1-k^2)^{3}
 \nonumber \\
 &\times (19+12{\cos}4\phi 
 +{\cos}8\phi ){\ln}(\frac{\sqrt{6(1-z^2)}e^{\gamma}
 \alpha }{\pi k_{B}T_c}) .
\end{align}
 Using  
\begin{equation}
\int _{0}^{1}dk_{z}(1-z^2)^3{\ln}(1-z^2)=
-\frac{2552}{3675}+\frac{16}{35}{\ln}4, 
\label{eqA28}
\end{equation}
we obtain from eq.~(\ref{eqA28})
\begin{equation}
 {\ln}(\frac{T_{c}}{T_{c0}})
 =
 -\frac{19}{20}\big( {\ln}(\frac{2\sqrt{6}e^{\gamma }\alpha }{\pi k_{B}T_c})\big)
 +\frac{19}{16} \cdot \frac{319}{525} ,
\end{equation}
which is written as
\begin{equation}
 {\ln}\left( (\frac{T_c}{T_{c0}})^{\frac{20}{19}}(\frac{\alpha }{\pi k_{B}T_c})\right)
 =
 {\ln}\big( \frac{e^{\frac{319}{400}-\gamma }}{2\sqrt{6}}\big) .
\end{equation}
Finally we obtain
\begin{align}
 \frac{T_c}{T_{c0}}
 &=\big( \frac{e^{\frac{319}{400}-\gamma }}{2\sqrt{6}}\big)^{19}
 \big[\frac{\alpha }{\pi k_{B}T_{c0}}\big] ^{-19} \nonumber \\
 &\approx\left( 0.254 \frac{\pi k_{B}T_{c0}}{\alpha}\right)^{19}\nonumber \\
 &\approx 5.08 \times 10^{-12}  \left(\frac{\pi k_{B}T_{c0}}{\alpha}\right) ^{19} .
\end{align}
The transition temperature is reduced as $\alpha^{-19}$ when $\alpha \gg 1$.

\end{document}